\documentclass[12pt]{article}
\usepackage{latexsym}
\usepackage{amsmath}
\usepackage{epsfig,graphics,psfrag}

\newcommand{\be}{\begin{equation}}
\newcommand{\ee}{\end{equation}}

\newlength{\figsize}
\begin{document}

\begin{titlepage}

\vspace*{0.7in}
 
\begin{center}
{\large\bf The deconfining phase transition in D=2+1 SU(N) gauge theories \\}
\vspace*{1.0in}
{Jack Liddle$^{a}$ and Michael Teper$^{b}$\\
\vspace*{.2in}
$^{a}$Fakultat fur Physik, Universitat Bielefeld, D-33615 Bielefeld,
Germany \\
\vspace*{.1in}
$^{b}$ Rudolf Peierls Centre for Theoretical Physics, University of Oxford,\\
1 Keble Road, Oxford OX1 3NP, UK\\
}
\end{center}

\vspace*{0.55in}

\begin{center}
{\bf Abstract}
\end{center}

We study the deconfining transition of SU($N$) gauge theories
in 2+1 dimensions for $2\leq N \leq 8$. We confirm that the
transition is second order for $N\leq 3$ and first 
order for $N\geq 5$. For the more delicate case of SU(4) all
our evidence points to the transition being weakly first order.
After extrapolating to the continuum limit, we obtain a 
deconfining temperature that can be well fitted by 
$T_c/\surd\sigma = 0.9026(23) + 0.880(43)/N^2$ for all $N\geq 2$.

\end{titlepage}

\setcounter{page}{1}
\newpage
\pagestyle{plain}

\section{Introduction}
\label{section_intro}

In 2+1 dimensions, SU($N$) gauge theories have a coupling $g^2$ 
that has dimensions of mass. Thus the dimensionless coupling
appropriate to physics on a length scale $l$ is
$g^2 l$: that is to say, the theory becomes free 
in the ultraviolet, and strongly coupled in the infrared.
Moreover the theory is found to be linearly confining
(see 
\cite{gs_k1}
and references therein) although no analytic proof exists.
These properties are reminiscent of $D=3+1$ gauge theories,
and so $D=2+1$ gauge theories have received much attention
in the hope that they might provide a useful stepping stone
between the soluble $D=1+1$ case, and the physically interesting
$D=3+1$ case. 

In this paper we calculate some properties of the deconfining
transition in $D=2+1$ SU($N$) gauge theories. We use standard
lattice methods, extrapolate to the continuum limit, and
then perform a large $N$ extrapolation that enables us to
extend our results to all values of $N$. These calculations
parallel similar calculations in $D=3+1$ 
\cite{d4_Tc}.

There exist older calculations for SU(2) and SU(3)
(see for example 
\cite{old_N2N3})
as well as more recent work on SU(4)
\cite{PdF_N4,jlmt_lat05,KH_N4}
and SU(5)
\cite{KH_N5,jlmt_lat05}
most of which appeared while this work was in progress.
Most of our calculations have appeared in
\cite{JL_Dphil}
which contains various details omitted in this paper,
and a summary of our early results was presented in
\cite{jlmt_lat05}.
The main change in this paper is that we use a more
accurate calculation of the string tension
\cite{gs_k1}
to provide the scale in which to express the quantities we
calculate, and we examine the delicate case of SU(4) 
in greater detail.

\section{Lattice setup}
\label{section_lattice}

Our calculations are performed using standard lattice tachniques. 
We work on periodic cubic $L^2_s\times L_t$ lattices with lattice 
spacing $a$. The degrees of freedom are SU($N$) matrices, $U_l$, 
assigned to the links $l$ of the lattice.
The partition function is 
\begin{equation}
Z(\beta)
=
\int \prod_l dU_l 
e^{-\beta \sum_p\{ 1 - \frac{1}{N}{\mathrm{ReTr}}U_p\}}
\label{eqn_lattice} 
\end{equation} 
where $U_p$ is the ordered product of matrices around the
boundary of the elementary square (plaquette) labelled by $p$.
In the continuum limit
\begin{equation}
\beta=\frac{2N}{a g^2}
\label{eqn_beta} 
\end{equation} 
where $g^2$ is the coupling, which in $D=2+1$ has dimensions
of mass. This is the standard Wilson plaquette action and 
the continuum limit is approached by tuning $a\to 0$ and
hence $\beta = 2N/ag^2 \to \infty$. One expects that for large $N$ 
physical masses will be proportional to the 't Hooft coupling 
$\lambda \equiv g^2 N$
\cite{largeN},
and this is indeed what one observes
\cite{glued3}.
So if we vary $\beta \propto N^2$ then we are keeping the lattice spacing 
$a$  fixed in physical units, to leading order in $N$.

When $L_s \to \infty$ at fixed $L_t$ and fixed $\beta$, the field
theory is at a temperature 
\begin{equation}
T= \frac{1}{a(\beta) L_t}.
\label{eqn_T} 
\end{equation} 
As is conventional, we use eqn(\ref{eqn_T}) to define $T$ at finite 
$L_s\gg L_t$ as well. By increasing $\beta$ at fixed $L_t$ we increase  
$T$ so as to locate the critical value $\beta=\beta_c$ at which the
system deconfines: $T_c(a)=1/a(\beta_c)L_t$. (We shall often write
this as $T_c$ if there is no ambiguity.)
An `order' parameter we shall use is the plaquette, $u_p$, averaged 
over the space time volume, 
\begin{equation} 
\overline{u}_p = \frac{1}{N_p} \sum_p u_p \quad ; \quad
u_p =  \frac{1}{N}{\mathrm{ReTr}}U_p 
\label{eqn_up} 
\end{equation} 
where $N_p=3L^2_sL_t$ is the total number of plaquettes. Another
parameter is the Polyakov loop
\begin{equation} 
l_p(x) = \mathrm{Tr} \prod^{L_t}_{t=1} U_t(x,t) \quad ; \quad
\overline{l}_p = \frac{1}{L_s^2} \sum_x l_p(x).
\label{eqn_poly} 
\end{equation} 
\section{Methodology}
\label{section_methods}

In this Section we briefly describe how we calculate the properties
of the deconfining phase transition. For a more detailed review, with
references, see for example 
\cite{HMO_phases}.

Consider the specific heat on a $L^2_sL_t$ lattice, defined 
conventionally by
\begin{equation} 
\frac{1}{\beta^2} C(\beta)
\equiv
\frac{\partial}{\partial\beta}\langle \overline{u}_p \rangle
=
N_p \langle \overline{u}^2_p \rangle
-
N_p \langle \overline{u}_p \rangle^2 .
\label{eqn_C} 
\end{equation} 
Suppose that there is a first order deconfining transition
at $\beta_c$ so that $\langle \overline{u}_p \rangle$
jumps from $\langle \overline{u}_{p,c} \rangle$ to
$\langle \overline{u}_{p,d} \rangle$ as we increase $\beta$
through $\beta_c$. On our finite spatial volume, $V$, the jump
becomes a crossover. Let us choose the maximum of $C(\beta)$
as defining the critical $\beta$ at finite volume: $\beta_c(V)$.
Up to the weakly varying factor of $\beta^{-2}$ this coincides
with the point at which  $\langle \overline{u}_p \rangle$
changes fastest with $\beta$. It is easy to see that for large
enough $V$ this is the value of $\beta$ at which the system
is equally likely to be in the confined and deconfined phases
and that
\begin{equation} 
\lim_{V\to\infty} \frac{1}{\beta_c^2} C(\beta_c(V))
=
\frac{N_p}{4}
\left(
\langle \overline{u}_{p,c} \rangle
-
\langle \overline{u}_{p,d} \rangle
\right)^2
\propto
V L^2_h.
\label{eqn_C1st} 
\end{equation} 
Thus at a first order transition the specific heat diverges
linearly in $V$ and the limiting value of $C(\beta)/V$
is directly related to the latent heat $L_h$.
Now, the crossover exists over a region of $T$ where there
is a finite probability for the system to be in both the
confined and deconfined phases. Assuming that the free
energies of the two phases are smooth as one passes through
$T_c$, this implies that the crossover extends over a
range of $O(1/V)$ in $T$ and hence that any definition
of $T_c$ within this range will generically have a leading $O(1/V)$
correction at large $V$. So we expect that as $V\to\infty$
\begin{equation} 
\frac{T_c(\infty)-T_c(V)}{T_c(\infty)}
=
\frac{\beta_c(\infty)-\beta_c(V)}{\beta_c(\infty)}
=
\frac{h}{VT_c^2}
=
h\left(\frac{L_t}{L_s}\right)^2
\label{eqn_h1st} 
\end{equation} 
(where we have used $T=1/aL_t=\beta g^2/2NL_t$). Moreover,
because the free energy $\propto N^2$, we also expect that
\begin{equation} 
h \stackrel{N\to\infty}{\propto} \frac{1}{N^2}
\label{eqn_hN} 
\end{equation} 
Thus if we calculate $\beta_c(V)$ from the maximum of $C(\beta)$, 
and do so for several $V$, we can use eqn(\ref{eqn_h1st})
to extrapolate to $V=\infty$, thus obtaining $\beta_c(V=\infty)$.
We can then use the results of
\cite{gs_k1}
to obtain the string tension at this $\beta$, and hence
$T_c(a)/\surd\sigma(a)$ at $a=1/L_tT_c$. We now repeat the exercise 
for several other values of $L_t$ and perform an extrapolation
of this ratio to the continuum limit. In the same way
we can use  eqn(\ref{eqn_C1st}) to obtain the continuum
value of the latent heat, e.g. $L_h/T^3_c$.

If the transition is second order, there is no discontinuity
in $\langle \overline{u}_p \rangle$ but there is a diverging
correlation length. Rewriting eqn(\ref{eqn_C}) as an
integrated correlation function
\begin{equation} 
\frac{1}{\beta^2} C(\beta)
=
\sum_p 
\langle 
\left(
\overline{u}_p - \langle\overline{u}_p\rangle
\right)
\left(
\overline{u}_{p_0} - \langle\overline{u}_p\rangle
\right)
\rangle
\label{eqn_Ccorr} 
\end{equation} 
(where $p_0$ is an arbitrary plaquette) we see that at the transition
$C(\beta)$ will diverge as $V\to\infty$, with a power
that depends on how the correlation length diverges with increasing
$V$, i.e. on the critical exponents. The maximum of
$C(\beta)$ is where the correlation length is largest
(up to weak corrections) and provides a natural
definition of $\beta_c(V)$. Using a conventional
notation, the leading large volume behaviour is
\begin{equation} 
\frac{1}{\beta^2} C(\beta)
=
c_0 V^{\frac{\alpha}{\nu d}} +c_1
\label{eqn_C2nd} 
\end{equation} 
and 
\begin{equation} 
\frac{T_c(\infty)-T_c(V)}{T_c(\infty)}
=
\frac{\beta_c(\infty)-\beta_c(V)}{\beta_c(\infty)}
=
\frac{h}{(VT_c^2)^{\frac{1}{\nu d}}}
=
h\left(\frac{L_t}{L_s}\right)^{\frac{2}{\nu d}}
\label{eqn_h2nd} 
\end{equation} 
We can use this to obtain $\beta_c(\infty)$ from our finite $V$
calculations, then use
\cite{gs_k1} 
to transform this into a value of $T_c(a)/\surd\sigma(a)$, then
repeat all this for several values of $L_t$, and finally extrapolate 
to the continuum limit. 

In practice using the plaquette and the specific heat is
not the most accurate method. For our second order transitions the 
excitation whose mass vanishes typically has a weak overlap with
the plaquette, so that it contributes a small peak to
$C(\beta)$ unless the volume because extremely large.
For our first order transitions the discontinuity is on
physical length scales, so  
$\left(
\langle \overline{u}_{p,c} \rangle
-
\langle \overline{u}_{p,d} \rangle
\right)  \sim O(a^3)$
and once again one has to go to very large $V$ to 
see a strong effect in $C(\beta)$. 

For these reasons it is usual to also use the
susceptibility of the Polyakov loop
\begin{equation} 
\frac{\chi}{V}
=
\langle |\overline{l}_p|^2 \rangle
-
\langle |\overline{l}_p| \rangle^2 .
\label{eqn_chi} 
\end{equation} 
In the confined phase the Polyakov loop fluctuates about zero
while in the deconfined phase it fluctuates about a non-zero
value proportional to an element of the center of SU($N$).
One uses $|\overline{l}_p|$ rather than  $\overline{l}_p$
since even in the deconfined phase the latter will (strictly speaking)
average to zero for finite $V$ through tunnelling between the 
$N$ (Euclidean) vacua. While taking the modulus might seem like an 
ad hoc fix, in fact it is closely related to the adjoint Polyakov loop: 
$\mathrm{Tr_A}l_p \propto |\mathrm{Tr_F}l_p|^2 -  1$.
The latter is a more natural and general `order' parameter than the 
usual fundamental Polyakov loop. For example, it will serve
equally well if we decide to express the gauge fields in
the adjoint reperesentation. For large $N$ it vanishes as $1/N^2$
in the confined phase, and is unsuppressed in the deconfined phase.
For these reasons it would make sense to use a susceptibility
based on the adjoint Polyakov loop. However, given that the definition in 
eqn(\ref{eqn_chi}) is now customary, we shall adhere to it in this paper. 
Just as for $C(\beta)$, we can define a $\beta_c(V)$ from the maximum 
of $\chi/V$ and extrapolate it to $V=\infty$ using eqn(\ref{eqn_h1st})
or eqn(\ref{eqn_h2nd}) depending on the order of the transition.
The value of $\chi/V$ at its maximum is not directly related to 
a physical quantity, unlike $C(\beta)$, so we do not consider
its infinite volume extrapolation. 

The reason that $\chi/V$ provides a better signal for the phase
transition than $C(\beta)$ is that the ground state of the flux loop 
that winds around the temporal torus becomes massless as 
$T\to T^-_c$ if the transition is second order. (This mass
is an eigenstate of a transfer matrix along a spatial direction.)
The Polyakov loop is an operator with a projection on this state.
The fact that near $T_c$ excited flux loops are relatively massive,
and that the projection on the ground state is quite good, for the 
range of $a$ we work with, means that the effect of the vanishing 
mass on $\chi/V$ is very pronounced. For a first order transition, 
on the other hand, the discontinuity  in the average Polyakov loop 
is not directly related to a physical energy density and so is not
as small as for the plaquette. As $\beta\to\infty$ Polyakov loops
do get renormalised to zero, and using  $\chi/V$ becomes more
difficult, but for our range $a \in [1/2T_c,1/5T_c]$ these problems
are not severe and accurate calculations are easy to perform.

We therefore  base our calculations of $\beta_c$ on  $\chi/V$. However
we do use $C(\beta)$ to obtain the latent heat in those cases
where the extrapolation in eqn(\ref{eqn_C1st}) can be usefully
performed. To do so we recall that the energy density is
given by
\begin{equation} 
\frac{\overline{E}}{V}
=
-\frac{1}{V}\frac{\partial}{\partial\frac{1}{T}}\ln Z
=
\frac{T^2}{V}\frac{\partial}{\partial T}\ln Z ,
\label{eqn_avEcont} 
\end{equation} 
where we use $Z=\sum_s \exp\{-E_s/T\}$, while 
\begin{equation} 
N_p \langle u_p \rangle
=
\frac{\partial}{\partial\beta}\ln Z 
=
\frac{\partial T}{\partial\beta}\frac{\partial}{\partial T}\ln Z 
+
\frac{\partial V}{\partial\beta}\frac{\partial}{\partial V}\ln Z 
\label{eqn_avElat} 
\end{equation} 
where the first equality in eqn(\ref{eqn_avElat})
uses the expression in eqn(\ref{eqn_lattice})
for $Z$. Noting that the volume derivative gives the pressure,
and that the pressure of the two phases is equal at $T=T_c$,
we see that the contribution of this second term to
\begin{equation} 
\Delta_{u_p} 
\equiv
\langle {u}_{p,c}\rangle - \langle {u}_{p,d} \rangle
\equiv
\langle \overline{u}_{p,c}\rangle - \langle\overline{u}_{p,d} \rangle
\label{eqn_Dup} 
\end{equation} 
will vanish at $T=T_c$. Thus the latent heat, $L_h$, which is the 
difference of the energy densities in the two phases at $T=T_c$, 
can be written, using 
eqns(\ref{eqn_avEcont},\ref{eqn_avElat},\ref{eqn_Dup}) and $T=1/aL_t$, as
\begin{equation} 
\frac{L_h}{T^3_c}
=
-3L^3_t a \frac{\partial\beta}{\partial a} \Delta_{u_p}
\simeq
3L^3_t\beta \Delta_{u_p}
\label{eqn_Lh} 
\end{equation} 
where at the last step we use
the asymptotic expression $\beta=2N/ag^2$. As we see
from eqn(\ref{eqn_C1st}), the quantity  $\Delta_{u_p}$,
and hence the latent heat, is given by the $V\to\infty$ limit 
of the specific heat. 

The expression in eqn(\ref{eqn_Lh}) is not unique. We can
do better by noting that $a\partial\beta/\partial a
= (a\mu)\partial\beta/\partial (a\mu)$ where we choose for $\mu$ 
some physical quantity whose lattice value has been
accurately calculated over a wide range of $\beta$
(e.g. the string tension). Now parameterise the variation
of $a\mu$ with $\beta$ as a series in $1/\beta$, evaluate
$(a\mu)\partial\beta/\partial (a\mu)$and substitute
in eqn(\ref{eqn_Lh}). This will improve over using $\beta=2N/ag^2$
by at least an additional $O(1/\beta)$ correction. Having said that,
we shall use the cruder estimate in eqn(\ref{eqn_Lh}) because
our relatively large errors on $L_h$ do not demand a more
sophisticated treatment.

For some $L_t$ our range of $V$ is not large enough to calculate
$L_h$ from the specific heat. In those case we take a sequence
of lattice fields generated close to $T=T_c$, and identify long
subsequences that are in either the confined or deconfined
phase, using the value of $\overline{l}_p$ to do so. We then obtain
$\langle {u}_{p,c} \rangle$ and $\langle {u}_{p,d} \rangle $
from the appropriate subsequences, and hence $\Delta_{u_p}$.
This direct method of calculating $\Delta_{u_p}$ has the
advantage of weak finite volume corrections and so does not 
require an explicit extrapolation to $V=\infty$. However
there are systematic errors to do with the assignment
of lattice fields to a particular phase, and these errors
only become small on large volumes where tunnellings
between phases become relatively rare. Finally, 
using eqn(\ref{eqn_Lh}) we transform our value of $\Delta_{u_p}$
into a value for the latent heat.  
 
An aside. Although one might imagine that
the calculations in $D=2+1$ would be easier than in $D=3+1$
this is not really so. For example, one requirement that a 
first order transition should be well defined in a practical
calculation is that the rate of tunnelling should be  small 
enough for the co-exisitence of the ordered and disordered
phases to be readily visible. This rate is governed by a 
factor  $\exp\{-2\sigma_{cd}L^{D-2}_s/T_c\} =
\exp\{-2\sigma^0 (L_s/L_t)^{D-2}\}$
where $\sigma_{cd}$ is the interface tension and we have written
$\sigma_{cd}=\sigma^0 T_c^{D-1}$. This factor gives
the suppression of those intermediate configurations in a
tunnelling where the volume is split into two co-existing phases
by domain walls that extend right across the lattice.
So all other things being equal, i.e. if the dimensionless tension  
$\sigma^0$ is the same, then the size $L_s/L_t$ in 2 space dimensions
should equal the square of the size, $(L_s/L_t)^2$, in 3 space
dimensions in order to achieve the same tunnelling suppression.
So the typical factor $L_s/L_t\sim 3$ in $D=3+1$ needs to
become a much more challenging  $L_s/L_t\sim 10$ in $D=2+1$.
For reasons like this, the accuracy and range of our calculations 
is not much more impressive than in $D=3+1$
\cite{d4_Tc}.
\section{Results}
\label{section_results}

\begin{figure}[htb]
\psfrag{chi}[c]{$\frac{\chi}{V}$}
\psfrag{beta}[c]{$\beta$}
\centerline{
\rotatebox{0}{
\includegraphics[width=11cm,height=8.5cm]{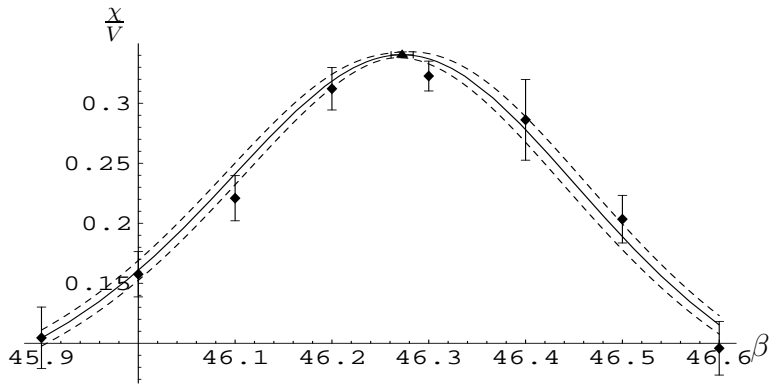}
}}
\caption{An example of reweighting: the Polyakov loop
susceptibility $\chi/V$ versus $\beta$ in SU(6) on a $25^2 3$ 
lattice.}
\label{fig_reweight}
\end{figure}

The basic step is to calculate the maximum of $\chi/V$, or $C(\beta)$,
for some given values of $N$, $L_t$ and $L_s$. To do so we typically
evaluate $\chi/V$ at 4 or more values of $\beta$. 
These values are judiciously chosen 
so as to straddle the maximum and to be close enough to that maximum 
for each Monte Carlo run to have a usefully large number of tunnelling
events between confined and deconfined phases. (All this clearly requires
some preparatory work.) To obtain the maximum accurately we need
a smooth interpolation between the calculated values, and this
is provided by standard reweighting techniques (see
\cite{reweight}
and Appendix A of
\cite{JL_Dphil}).
In Fig.~\ref{fig_reweight} we show an example of such a reweighted
curve, for a $25^2 3$ lattice in SU(6).

We now present our results. We begin with SU(2) and SU(3) where
the deconfinement transitiona are well known to be second order
\cite{old_N2N3}.
We then deal with SU($N\geq 5$) where they  turn out to be 
clearly first order. Finally we dwell in
more detail on the delicate case of SU(4). In each case we perform
calculations for $L_t=2,3,4,5$ and from these we obtain
the continuum limit of $T_c/\surd\sigma$ and (where relevant)
$L_h/T^3_c$. The continuum limit  can be obtained from a standard 
weak coupling extrapolation of the form
\begin{equation} 
\frac{aT_c(a)}{a\surd\sigma(a)}
= 
\frac{T_c(a)}{\surd\sigma(a)}
=
\frac{T_c(0)}{\surd\sigma(0)}
+ c_1 a^2\mu^2 + c_2 a^4\mu^4 + \cdots .
\label{eqn_cont} 
\end{equation} 
Here $\mu$ is some convenient physical energy scale, such as 
$\surd\sigma$ or $T_c$, in which case $a^2T^2_c = 1/L^2_t$. 
The string tension is obtained 
by interpolation of the results presented in 
\cite{gs_k1}.
In the final subsection we examine the $N$-dependence
of these quantities as well as the scaling of finite volume
corrections.

One might reasonably worry that at $L_t=2$ the lattice spacing is 
too coarse for the weak coupling expansion in eqn(\ref{eqn_cont})
to be applicable. (In units of $T_c$ it will be $a=0.5 T_c$.) 
In fact there is a strong-to-weak coupling crossover in $D=2+1$ 
SU($N$) gauge theories that becomes a phase transition at $N=\infty$
\cite{fbmt}
in a manner very similar to the Gross-Witten transition in $D=1+1$
\cite{GW}.
This transition occurs at a value of $\beta$ that is about $10\%$
below $\beta_c(L_t=2)$. Thus it is not excluded that at this $\beta$
a weak-coupling  expansion is applicable. We also note that for light
glueball masses the ratio $m_G/\surd\sigma$ appears to be described
by such a weak-coupling expansion at comparable values of $\beta$
\cite{glued3,glued3_other}.
For these reasons we shall (cautiously) attempt continuum
extrapolations from $L_t=2$.

\subsection{SU(2) and SU(3)}
\label{subsection_N2N3}

It is known from earlier calculations 
\cite{old_N2N3}
that both SU(2) and SU(3) have second order deconfining transitions, 
and there is good evidence that these are in the universality classes
of the $2D$ $q=2$ and $q=3$ state Potts models respectively,
as one would expect from the general arguments of
\cite{SY}.
Our calculations corroborate the second order nature of the
transitions. As an example we show in Fig.~\ref{fig_2ndhisto}
a histogram of values of $\overline{l}_p$ as obtained in SU(2) 
on a $90^2 3$ lattice at $\beta=4.9$, which is very close to
$\beta_c(L_t=3)$. We observe a single broad distribution
which shows no sign of the two peak structure that would
indicate co-existing phases.    

For each value of $L_t=1/aT_c$ in the range $L_t\in [2,5]$ we calculate 
$\beta_c(V)$ for several values of $V$ and extrapolate to $V=\infty$ 
using eqn(\ref{eqn_h2nd}) with $\nu=1,5/6$ for $N=2$ and $N=3$
respectively. (These being the exponents of the relevant Potts models
\cite{Potts}.) 
We show an example of such an extrapolation in
Fig.~\ref{fig_bcVsu2}; it is for SU(2) and $L_t=3$.
In Table~\ref{table_n2Tc} we list the values of $\beta_c(V=\infty)$ 
for SU(2) that have been obtained in this way, and the range
of volumes used in each extrapolation. We also give the
value of the string tension
\cite{gs_k1}
that corresponds to each $\beta_c(V=\infty)$ for each $L_t$.
(The error on $a\surd\sigma$ includes the error on $\beta_c$.)
From this we obtain the values of
$T_c/\surd\sigma = 1/a\sqrt{\sigma}L_t$ shown in
Table~\ref{table_n2Tc}. In Table~\ref{table_n3Tc}
we display the corresponding results for SU(3).

\subsection{SU($N\geq 5$)}
\label{subsection_N5to8}

For $N=5$ it is known that the transition is first order
\cite{KH_N5,jlmt_lat05}
and our calculations show that this is the case up to
$N=8$. We find that the strength of the transition is `normal' 
and that the latent heat is approximately constant for
$N\geq 5$. This provides convincing evidence for the
claim that the transition remains first order for 
larger $N$ and, in particular, for $N=\infty$.

\begin	{figure}[p]
\begin	{center}
\leavevmode
\input	{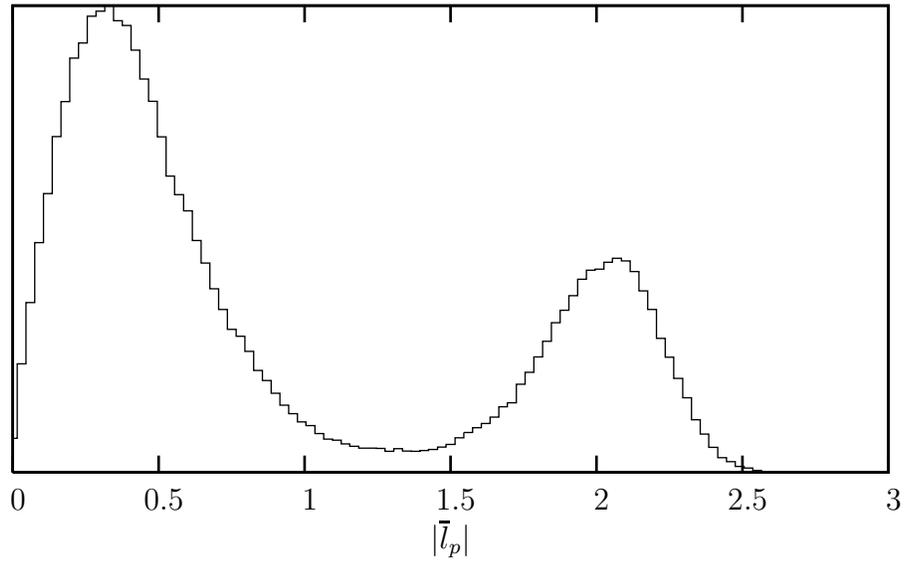}
\end	{center}
%\vskip 0.15in
\caption{Number of lattice fields, $N$, with a given
value of the Polyakov loop $|\overline{l}_p|$.
In SU(8), at $\beta=82.5$, on a $20^2 3$ lattice.}
\label{fig_1sthisto}
\end 	{figure}

To illustrate the first order nature of the transition
we show in Fig.~\ref{fig_1sthisto}
a histogram of values of $\overline{l}_p$ as obtained in SU(8) 
on a $20^2 3$ lattice at value of $\beta$ which is very close to
$\beta_c(L_t=3)$. We observe a pronounced two peak structure
that clearly indicates the co-existence of ordered and
disordered  phases. Moreover one finds that this two-peak
structure becomes more pronounced with increasing $V$.

For each value of $L_t=1/aT_c$ in the range $L_t\in [2,5]$ we calculate 
$\beta_c(V)$ for several values of $V$ and extrapolate to $V=\infty$ 
using eqn(\ref{eqn_h1st}).
We show an example of such an extrapolation in
Fig.~\ref{fig_bcVsu6}; it is for SU(6) and $L_t=3$.
In Tables~\ref{table_n5Tc},~\ref{table_n6Tc},~\ref{table_n8Tc} 
we list the values of $\beta_c(V=\infty)$ 
that have been obtained in this way, and the ranges
of volumes used in each extrapolation. We also give the
values of the string tension
\cite{gs_k1}
that correspond to each $\beta_c(V=\infty)$ for each $L_t$.
From this we obtain the values of
$T_c/\surd\sigma = 1/a\sqrt{\sigma}L_t$ shown in the Tables.

In Table~\ref{table_hTc} we list our fitted value of the coefficient
$h$ of the $O(1/V)$ correction to $\beta_c(V)$ in
eqn(\ref{eqn_h1st}).

To obtain the specific heat, $L_h$, as given by eqn(\ref{eqn_Lh}),
we calculate the quantity
$\Delta_{u_p}$ defined in eqn(\ref{eqn_Dup}). For small $L_t$ we
can obtain it from the specific heat peak as in eqn(\ref{eqn_C1st}).
For larger $L_t$ this becomes inaccurate and we obtain it
by identifying sequences of confined and deconfined lattice fields,
as described at the end of Section~\ref{section_methods}. Our
results are listed in Table~\ref{table_Delup}. Values obtained
from the specific heat are labelled by a star. In most cases
where we have values from both methods, they are reasonably
consistent. In one case this is clearly not so: this serves
as a warning that, at least for this rather delicate quantity, 
our claimed control of systematic errors is likely to be
optimistic.

\subsection{SU(4)}
\label{subsection_N4}

If the SU(4) transition were second order, one would expect
\cite{SY,PdF_N4}
it to be in the universality class of the symmetric Ashkin-Teller 
model (see
\cite{Sokal_AT}
for a recent review) of which the $q=4$ Potts model is a special 
case. This has continuously varying critical exponents and
logarithmic effects which, as emphasised in
\cite{PdF_N4},
creates difficulties in the analysis. The results of
\cite{PdF_N4},
favoured the Potts model class, although their $L_t=2$
transition appeared to be first order. Our preliminary results
\cite{jlmt_lat05}
pointed to the transition remaining first order at larger
$L_t$ and this was confirmed by our more extensive
calculations in
\cite{JL_Dphil}.
A simultaneous recent study of this question presents quite 
convincing evidence that the transition is first order
\cite{KH_N5}.
In this paper we present some of our results from
\cite{JL_Dphil}
as well as some from larger volumes.

We begin by showing the histogram of the Polyakov loop
on very large volumes, at our smallest values of the lattice
spacing. In Fig.~\ref{fig_histN4L60} we show the results 
obtained on a $60^2 5$ lattice and in Fig.~\ref{fig_histN4L120} 
those obtained on a $120^2 5$
lattice. We see a clear 2-peak structure reflecting the
co-existence of ordered and disordered phases. The structure
becomes more pronounced as $L_s/L_t$ increases from 12 to 24
and $V$ increases by a factor of 4. The gap between the two
maxima does not appear to shrink significantly as $V$
increases. All this provides rather strong evidence that the
transition is first order and remains so in the continuum limit.

In Fig.~\ref{fig_bcVsu4} we show an extrapolation of 
$\beta_c(V)$ assuming an $O(1/V)$ correction that assumes
it is indeed first order. As we see, such first order
finite size scaling works very well. In Table~\ref{table_n4Tc}
we list our extrapolated values of $\beta_c(V=\infty)$,
and the corresponding values of $T_c/\surd\sigma$.
In Table~\ref{table_hTc} we list the coefficient of
the $O(1/V)$ correction in eqn(\ref{eqn_h1st}). Finally
in  Table~\ref{table_Delup} we list the values of
the discontinuity in the average plaquette across the transition.
These will translate into values of the specific heat
through eqn(\ref{eqn_Lh}).

\subsection{continuum limits}
\label{subsection_cont}

To extrapolate our calculated  values of $T_c/\surd\sigma$ 
to the continuum limit we use eqn(\ref{eqn_cont}) with 
$a\mu = a\surd\sigma$. We could equally well choose to
use $a\mu = aT_c = 1/L_t$, and we have checked that the
results of the two types of fits are within one standard
deviation of each other. (We also find that the quality
of the fits is comparable.) In Table~\ref{table_contTc} we
show the results of two such fits: one is an $O(a^4)$
extrapolation to $L_t \geq 2$ and the second is an
$O(a^2)$ extrapolation to $L_t \geq 3$. One finds
that $O(a^2)$ extrapolations to $L_t \geq 2$ are
statistically unacceptable. 

\begin	{figure}[p]
\begin	{center}
\leavevmode
\input	{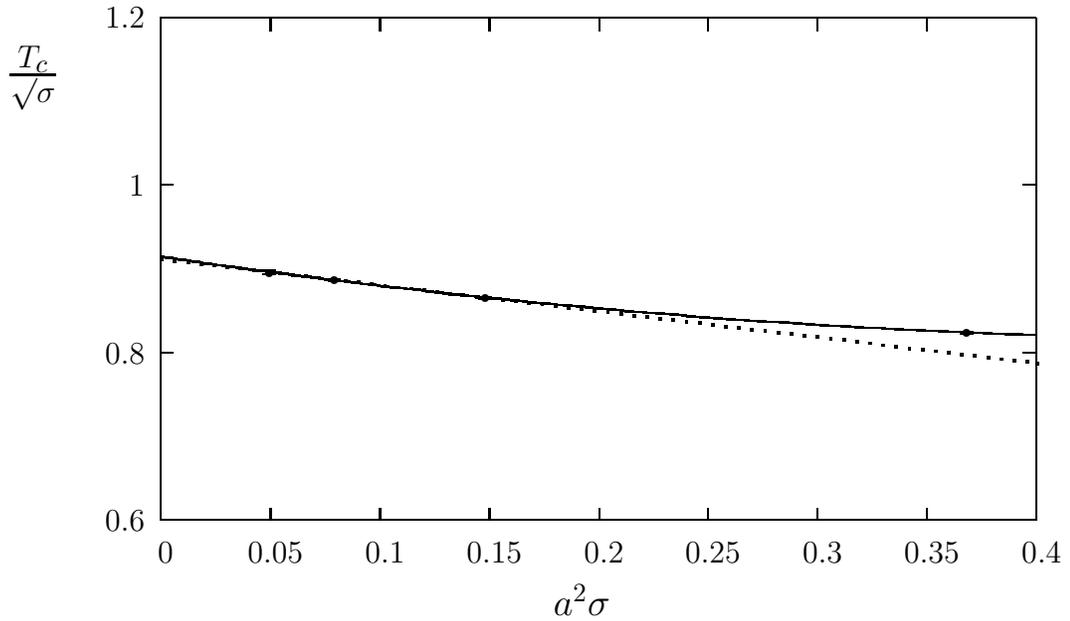}
\end	{center}
%\vskip 0.15in
\caption{Values in SU(8) of $T_c(a)/\surd\sigma(a)$ plotted against
$a^2\sigma$ and extrapolated to the continuum limit using
either linear (...) or quadratic (-) extrapolations.}
\label{fig_TKN8}
\end 	{figure}

The best fits are rather poor, but not unacceptable, for 
SU(2) and SU(3) and this presumably reflects the presence
of a very smooth strong-weak coupling crossover
at these smaller values of $N$. For SU(5) the fit
is extremely poor, and this is due entirely to our
$L_t=5$ value. As we see in Table~\ref{table_hTc}
the fitted coefficient of the $O(1/V)$ correction
is anomalously large. It seems clear that this
value happens to be out by several standard deviations.
We therefore also show a fit (denoted by a $\star$) that
uses the values of $\beta_c(V=\infty)$ given in   
\cite{KH_N5}
for $L_t=3,4,5$ together with our value for $L_t=2$
(since this is not calculated in
\cite{KH_N5}).
This gives a fit with an acceptable $\chi^2$. In fact
our values for $L_t=3,4$ are consistent with those of
\cite{KH_N5};
it is only our $L_t=5$ value that is anomalous.
This is reflected in the fact that the continuum
extrapolation of our results is very similar to that
obtained using the values of 
\cite{KH_N5},
as we see in Table~\ref{table_contTc}. What we can therefore
say is that for larger $N$ one has good $O(a^4)$ continuum
extrapolations even from $L_t=2$. As an example, we display
the SU(8) case in Fig.~\ref{fig_TKN8}. The residual
systematic error reflected in the small difference between
the $O(a^2)$ and  $O(a^4)$ fits would be removed by
calculations at slightly larger $L_t$.

Turning now to the latent heat, we find that we can
obtain good continuum extrapolations from $L_t\geq 2$
using just a $O(a^2)$ correction:
\begin{equation} 
\frac{1}{T_c(a)}\left(\frac{L_h(a)}{N^2}\right)^\frac{1}{3}
= 
\frac{1}{T_c(0)}\left(\frac{L_h(0)}{N^2}\right)^\frac{1}{3}
+ c_1 a^2\sigma .
\label{eqn_Lhcont} 
\end{equation} 
We normalise by $T_c$ because that is a characteristic
energy scale of the transition, and by $N^2$ because
one expects $L_h \propto N^2$ at large $N$. We choose
to take the third root so that we extrapolate a dimensionless
ratio that is of the form $[m]/[m]$. The results are
shown in Table~\ref{table_LhTc}. 

The values of the coefficient $h$ at larger values of $L_t$
are too inaccurate, as we see from Table~\ref{table_hTc} to allow 
an attempt at a useful continuum extrapolation of that quantity.

\subsection{$N$ dependence}
\label{subsection_N}

\begin	{figure}[p]
\begin	{center}
\leavevmode
\input	{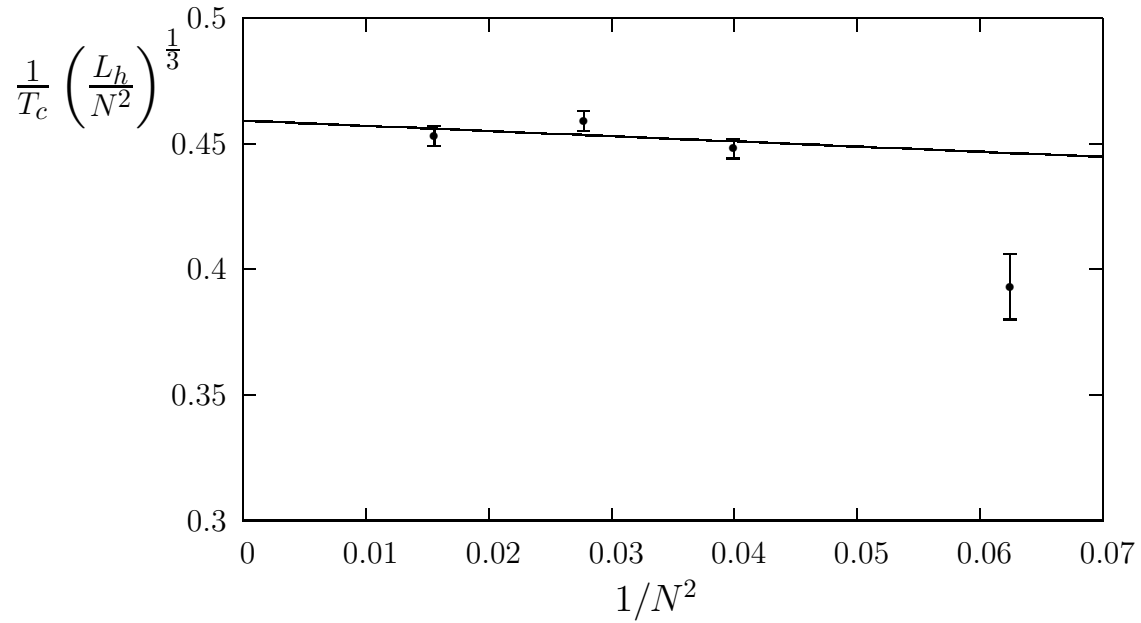}
\end	{center}
%\vskip 0.15in
\caption{Latent heat, with a natural normalisation, obtained
at a fixed lattice spacing $a=1/3T_c$. The fit to large $N$
uses a leading $O(1/N^2)$ correction.}
\label{fig_LhTN}
\end 	{figure}

In Fig.~\ref{fig_LhTN} we plot the continuum latent heat,
in units of $T_c$ and normalised to $N^2$, against 
$1/N^2$, which is expected to be the leading correction
to the $N=\infty$ limit. We obtain a good fit to $N\geq 5$ with
\begin{equation} 
\frac{1}{T_c}\left(\frac{L_h}{N^2}\right)^\frac{1}{3}
= 
0.459(7) - \frac{0.21(24)}{N^2}.
\label{eqn_LhN} 
\end{equation} 
It is not possible to include the SU(4) value.
Indeed in our natural units, the SU(4) latent heat $L_h$ is about
one-half of its value at larger $N$, where the variation with
$N$ is essentially negligible. That is to say, the deconfining
transition in SU(4) is weakly first order.

\begin	{figure}[p]
\begin	{center}
\leavevmode
\input	{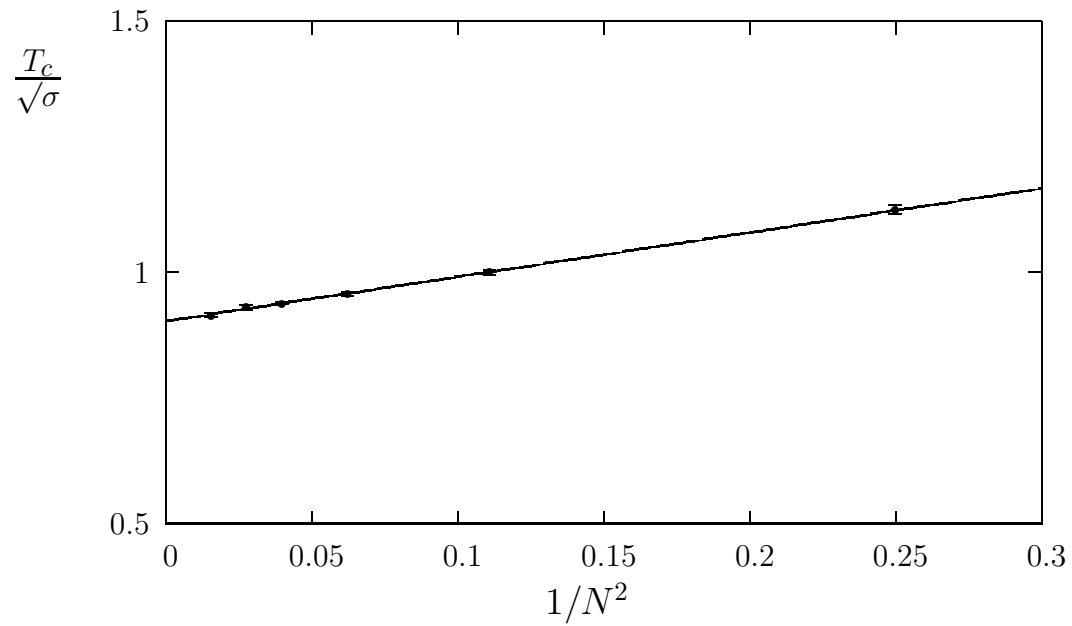}
\end	{center}
%\vskip 0.15in
\caption{Continuum values of $T_c/\surd\sigma$ plotted 
against $1/N^2$. The fit to all $N$ uses just a leading 
$O(1/N^2)$ correction.}
\label{fig_TKN}
\end 	{figure}

The result of attempting to fit our continuum values of 
$T_c/\surd\sigma$ with just a leading $O(1/N^2)$ correction
is shown in Fig.~\ref{fig_TKN}. (We use the values obtained 
by $O(a^4)$ extrapolations from $L_t\geq 2$.)
It is remarkable that one obtains a good fit
\begin{equation} 
\frac{T_c}{\surd\sigma}
=
0.9026(23) + \frac{0.880(43)}{N^2} 
\label{eqn_TcN} 
\end{equation} 
to all values of $N\geq 2$ despite the fact that the $N=2,3$
transitions are second order and the remainder are first order.

\begin	{figure}[p]
\begin	{center}
\leavevmode
\input	{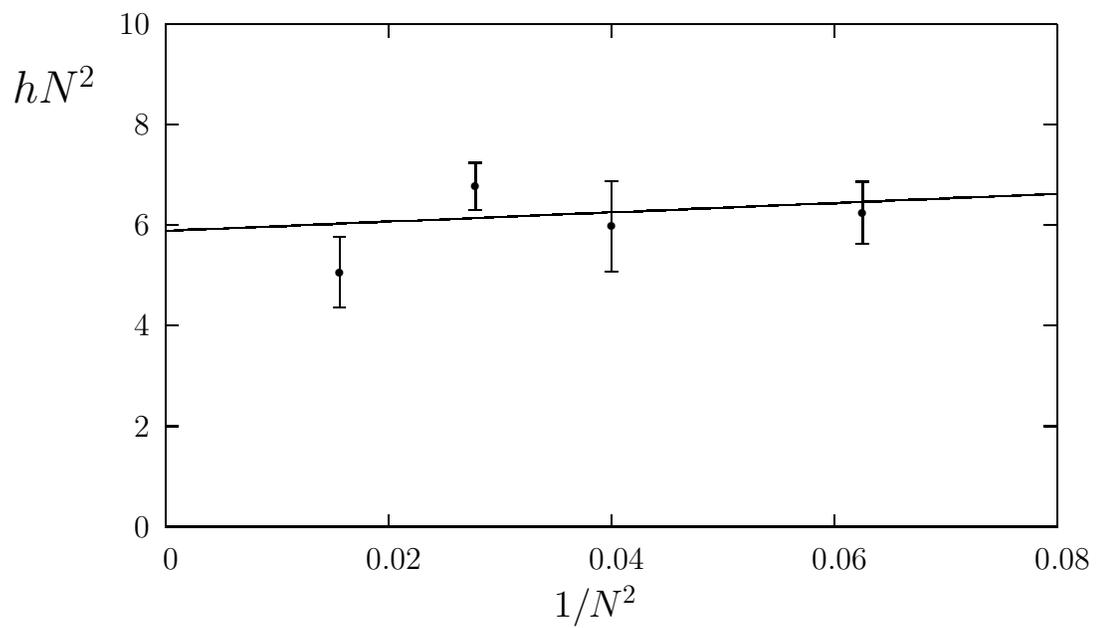}
\end	{center}
%\vskip 0.15in
\caption{$N$-dependence of the coefficient $h$  of the $O(1/V)$ 
correction to $\beta_c(V)$, obtained at a fixed lattice
spacing $a=1/3T_c$.}
\label{fig_hN}
\end 	{figure}

It is interesting to see if the finite volume effects
disappear with $N^2$, as one expects if the free energies and
interface tensions scale as $\propto N^2$. A relevant
quantitiy is the parameter $h$ that is the coefficient of the
$O(1/V)$ correction to the $V$-dependence of $T_c$. 
As we see from Table~\ref{table_hTc} the errors are too large 
at larger $L_t$ to allow a continuum extrapolation. Instead we
take the values at $L_t=3$, i.e. $aT=1/3T_c$, where the errors
are still small and $a$ is perhaps not too coarse. Plotting
the corresponding values of $hN^2$ in Fig.~\ref{fig_hN}
we see very nice evidence that finite volume corrections
vanish as $h\propto 1/N^2$ just as counting arguments suggest.

\section{Conclusions}
\label{section_conc}

We have found that, just as in $D=3+1$
\cite{d4_Tc},
SU($N$) gauge theories in $D=2+1$ have
a deconfining transition that is second order at low $N$
and first order at larger $N$. Again as in $D=3+1$
\cite{d4_Tc},
one finds that the variation with $N$ of $T_c/\surd\sigma$
is small and that, quite remarkably, one can fit this ratio 
for all $N\geq 2$ with just a leading $O(1/N^2)$ correction 
with a modest coefficient, as we see in Fig.~\ref{fig_TKN}.

We have seen that the SU(4) transition is almost certainly first 
order, although weakly so, as we see from its suppressed latent 
heat in Fig.~\ref{fig_LhTN}. This is reminiscent of $D=3+1$
where it is the SU(3) transition that is weakly first order.

When expressed in a natural normalisation, the value of
the latent heat at larger $N$ is 
$L_h \sim N^2 \times (0.46 T_c)^3$. By contrast, in
$D=3+1$ it is considerably  larger
\cite{d4_Tc}:
$L_h \sim N^2 \times (0.8 T_c)^4$. This is perhaps
natural given that the thermal gluons at high $T$ have
more degrees of freedom in three space dimensions than 
in two. But one should also note that if we express
$L_h$ in units of $\surd\sigma$ rather than $T_c$, most
of this difference goes away.

The fact that the value of  $T_c/\surd\sigma$ is $\sim 1.5$ 
times larger in $D=2+1$ than in $D=3+1$ is quite natural,
and already occurs for the simple Nambu-Goto string model
where the deconfining (Hagedorn) transition occurs
at $T_c^2 = 3/\pi(D-2)$
\cite{gs_k1,bbmt_H}.
This simply reflects the differing 
entropy of strings or flux tubes in 2 and 3 spatial dimensions.

It is obviously  interesting to ask if the gluon plasma 
immediately above the phase transition is strongly
coupled, as it has been found to be in $D=3+1$ both
experimentally and theoretically. (See
\cite{bbmt_SGP}
and references therein.) Here the effective
dimensionless  't Hooft coupling on the scale $T_c$ will be
\begin{equation} 
\frac{g^2N}{T_c}
\sim
\frac{g^2N}{\surd\sigma}
\sim
5
\end{equation} 
using the values in 
\cite{gs_k1}.
This is large enough to raise the possibility of a `strongly
coupled gluon plasma' in $D=2+1$ and  to motivate a numerical 
exploration, but that goes beyond the scope of the present paper.

\section*{Acknowledgements}

We have had useful discussions with B. Bringoltz and 
F. Bursa during the course of this work. We are grateful 
to Ph. de Forcrand for
critical discussions on the SU(4) transition and for
encouraging us to strengthen our numerical results
for that case.
The computations were performed on resources funded
by Oxford and EPSRC.  Jack Liddle's contribution was performed at Oxford and was supported by PPARC.

\vfill\eject

\begin{table}
\begin{center}
\begin{tabular}{|c|c|c|c|c|}\hline
\multicolumn{5}{|c|}{ SU(2) } \\ \hline
$aT_c$ &  $L_s\in$ &  $\beta_c$ & $a\surd\sigma$  & $T_c/\surd\sigma$  \\ \hline
\rule[-1mm]{0mm}{6mm}$\frac{1}{2}$ & [15,60] & 3.4475(36) & 0.4902(13) & 1.0200(27) \\
\rule[-1mm]{0mm}{6mm}$\frac{1}{3}$ & [22,90] & 4.943(13)  & 0.3164(10) & 1.0535(33) \\
\rule[-1mm]{0mm}{6mm}$\frac{1}{4}$ & [30,80] & 6.483(26)  & 0.2317(11) & 1.0790(51) \\
\rule[-3mm]{0mm}{8mm}$\frac{1}{5}$ & [50,85] & 8.143(57)  & 0.1799(14) & 1.1117(87) \\ \hline
\end{tabular}
\caption{\label{table_n2Tc} Critical $\beta$ at various values of
$L_t=1/aT_c$, with resulting values of $T_c/\surd\sigma$. String
tensions from \cite{gs_k1}. All for SU(2).}
\end{center}
\end{table}

\begin{table}
\begin{center}
\begin{tabular}{|c|c|c|c|c|}\hline
\multicolumn{5}{|c|}{ SU(3) } \\ \hline
$aT_c$ &  $L_s\in$ &  $\beta_c$ & $a\surd\sigma$  & $T_c/\surd\sigma$  \\ \hline
\rule[-1mm]{0mm}{6mm}$\frac{1}{2}$ & [15,60]  & 8.1489(31)  & 0.5641(17)  & 0.8864(27) \\
\rule[-1mm]{0mm}{6mm}$\frac{1}{3}$ & [23,90]  & 11.3711(91) & 0.35730(54) & 0.9329(14) \\
\rule[-1mm]{0mm}{6mm}$\frac{1}{4}$ & [30,120] & 14.717(17)  & 0.26098(41) & 0.9579(15) \\
\rule[-3mm]{0mm}{8mm}$\frac{1}{5}$ & [40,70]  & 18.131(61)  & 0.20517(79) & 0.9748(38) \\ \hline
\end{tabular}
\caption{\label{table_n3Tc} As in Table~\ref{table_n2Tc} but for SU(3).}
\end{center}
\end{table}

\begin{table}
\begin{center}
\begin{tabular}{|c|c|c|c|c|}\hline
\multicolumn{5}{|c|}{ SU(4) } \\ \hline
$aT_c$ &  $L_s\in$ &  $\beta_c$ & $a\surd\sigma$  & $T_c/\surd\sigma$  \\ \hline
\rule[-1mm]{0mm}{6mm}$\frac{1}{2}$ & [13,30]  & 14.8403(26)  & 0.5840(38) & 0.8562(56) \\
\rule[-1mm]{0mm}{6mm}$\frac{1}{3}$ & [20,45]  & 20.377(11)  & 0.37254(52) & 0.8948(13) \\
\rule[-1mm]{0mm}{6mm}$\frac{1}{4}$ & [30,60]  & 26.228(75)  & 0.27195(95) & 0.9193(32) \\
\rule[-3mm]{0mm}{8mm}$\frac{1}{5}$ & [37,60]  & 32.154(71)  & 0.21423(56) & 0.9336(25) \\ \hline
\end{tabular}
\caption{\label{table_n4Tc} As in Table~\ref{table_n2Tc} but for SU(4).}
\end{center}
\end{table}

\begin{table}
\begin{center}
\begin{tabular}{|c|c|c|c|c|}\hline
\multicolumn{5}{|c|}{ SU(5) } \\ \hline
$aT_c$ &  $L_s\in$ &  $\beta_c$ & $a\surd\sigma$  & $T_c/\surd\sigma$  \\ \hline
\rule[-1mm]{0mm}{6mm}$\frac{1}{2}$ & [12,20]  & 23.5315(58) & 0.5973(19)   & 0.8371(27) \\
\rule[-1mm]{0mm}{6mm}$\frac{1}{3}$ & [18,25]  & 32.065(21)  & 0.37912(58)  & 0.8792(14) \\
\rule[-1mm]{0mm}{6mm}$\frac{1}{3}$ &     & $32.0765(54)^\star$ & 0.37888(52) & 0.8798(12) \\ 
\rule[-1mm]{0mm}{6mm}$\frac{1}{4}$ & [24,36]  & 41.057(36)  & 0.27729(40)  & 0.9016(13) \\
\rule[-1mm]{0mm}{6mm}$\frac{1}{4}$ &     & $41.113(12)^\star$ & 0.27682(30)  & 0.9031(10) \\ 
\rule[-1mm]{0mm}{6mm}$\frac{1}{5}$ & [30,40]  & 50.67(16)   & 0.21626(80)  & 0.9248(34) \\ 
\rule[-3mm]{0mm}{8mm}$\frac{1}{5}$ &     & $50.275(20)^\star$ & 0.21823(22)  & 0.9165(10) \\ \hline
\end{tabular}
\caption{\label{table_n5Tc} As in Table~\ref{table_n2Tc} but for SU(5).}
\end{center}
\end{table}

\begin{table}
\begin{center}
\begin{tabular}{|c|c|c|c|c|}\hline
\multicolumn{5}{|c|}{ SU(6) } \\ \hline
$aT_c$ &  $L_s\in$ &  $\beta_c$ & $a\surd\sigma$  & $T_c/\surd\sigma$  \\ \hline
\rule[-1mm]{0mm}{6mm}$\frac{1}{2}$ & [10,20] & 34.1848(52) &
0.6033(35) & 0.8288(48)  \\
\rule[-1mm]{0mm}{6mm}$\frac{1}{3}$ & [15,25] & 46.395(17) &
 0.38058(94) & 0.8759(22) \\
\rule[-1mm]{0mm}{6mm}$\frac{1}{4}$ & [20,30] & 59.372(48) &
 0.27883(46) & 0.8966(15) \\
\rule[-3mm]{0mm}{8mm}$\frac{1}{5}$ & [30,37] & 72.59(24)  &
0.21975(88) &  0.9101(36) \\ \hline
\end{tabular}
\caption{\label{table_n6Tc} As in Table~\ref{table_n2Tc} but for SU(6).}
\end{center}
\end{table}

\begin{table}
\begin{center}
\begin{tabular}{|c|c|c|c|c|}\hline
\multicolumn{5}{|c|}{ SU(8) } \\ \hline
$aT_c$ &  $L_s\in$ &  $\beta_c$ & $a\surd\sigma$  & $T_c/\surd\sigma$  \\ \hline
\rule[-1mm]{0mm}{6mm}$\frac{1}{2}$ & [5,14]  & 61.245(11) & 0.60688(92) & 0.8239(13) \\
\rule[-1mm]{0mm}{6mm}$\frac{1}{3}$ & [10,20] & 82.704(43) & 0.38525(54) & 0.8652(12) \\
\rule[-1mm]{0mm}{6mm}$\frac{1}{4}$ & [10,18] & 105.52(11) & 0.28190(43) & 0.8868(14) \\
\rule[-3mm]{0mm}{8mm}$\frac{1}{5}$ & [12,25] & 128.30(28) & 0.22332(61) & 0.8956(25) \\ \hline
\end{tabular}
\caption{\label{table_n8Tc} As in Table~\ref{table_n2Tc} but for SU(8).}
\end{center}
\end{table}

\begin{table}
\begin{center}
\begin{tabular}{|c|c|c|c|c|}\hline
\multicolumn{5}{|c|}{$h$ } \\ \hline
$aT_c$ & SU(4) &  SU(5) & SU(6) & SU(8) \\  \hline
\rule[-1mm]{0mm}{6mm}$\frac{1}{2}$ & 0.275(14) & 0.190(15) & 0.132(7)  & 0.0567(20) \\ 
\rule[-1mm]{0mm}{6mm}$\frac{1}{3}$ & 0.390(39) & 0.239(36) & 0.188(13) & 0.079(11) \\
\rule[-1mm]{0mm}{6mm}$\frac{1}{4}$ & 0.50(19)  & 0.212(50) & 0.204(29) & 0.066(16) \\
\rule[-3mm]{0mm}{8mm}$\frac{1}{5}$ & 0.88(24)  & 0.71(15)  & 0.25(14)  & 0.052(29) \\ \hline
\end{tabular}
\caption{\label{table_hTc} The coefficient of the $O(1/V)$ term in
the variation of $\beta_c(V)$ in eqn(\ref{eqn_h1st}).}
\end{center}
\end{table}

\begin{table}
\begin{center}
\begin{tabular}{|c|c|c|c|c|}\hline
\multicolumn{5}{|c|}{$\Delta_{u_p}$ at $T=T_c$} \\ \hline
$L_t$ & SU(4) &  SU(5) & SU(6) & SU(8) \\  \hline
 2 & $0.00603(9)^\star$ & $0.00886(11)^\star$  & $0.01118(10)^\star$  &  $0.01238(9)^\star$   \\
   & $0.00631(37)^\dagger$      &  &  &    \\ \hline
 3 & $0.00073(8)^{\star\dagger}$  & $0.00117(12)^\star$  & $0.00138(9)^{\star\dagger}$  &  $0.00206(6)^\star$  \\
   & 0.00087(6)  & $0.00125(4)^\dagger$     &        &   $0.00146(3)^\dagger$  \\  \hline
 4 & 0.00024(5)  & 0.000336(9)     & 0.000352(20)  &  0.000377(24)  \\
\hline
 5 & 0.00010(1)  & 0.000140(5)     & 0.000148(4)   &  0.000157(8)  \\ \hline
\end{tabular}
\caption{\label{table_Delup} The difference between the average
plaquette in the confined and deconfined phases at $T=T_c$. Entries
with a $\star$ are calculated from the specific heat. Where there are 
two values, the $\dagger$ indicates the value used in the continuum 
fit of the latent heat.}
\end{center}
\end{table}

\begin{table}
\begin{center}
\begin{tabular}{|c|cccc|ccc|}\hline
\multicolumn{8}{|c|}{ continuum limit} \\ \hline
\multicolumn{1}{|c|}{} & \multicolumn{4}{c|}{$O(a^4)$ fit $L_t\geq 2$} & \multicolumn{3}{c|}{$O(a^2)$ fit $L_t\geq 3$}\\ \hline
$N$ &  $T_c/\surd\sigma$ &  $c_1$ & $c_2$  & $\chi^2/n_{df}$  &
$T_c/\surd\sigma$ &  $c_1$ & $\chi^2/n_{df}$  \\ \hline
 2 & 1.1238(88) & -0.87(14) & 1.83(45) & 3.3 & 1.1224(90) & -0.70(15)
& 3.3  \\
 3 & 0.9994(40) & -0.65(6) & 0.94(15) & 2.1  & 0.9890(31) & -0.44(4) &
1.9 \\
 4 & 0.9572(39) & -0.56(5) & 0.76(13) & 0.1  & 0.9516(33) & -0.41(3) &
0.5 \\
 5 & 0.9404(38) & -0.53(5) & 0.68(11) & 8.2 & 0.9325(28) & -0.38(3) & 11.1 \\
 $5^\star$ & 0.9380(19) & -0.49(3) & 0.58(8) & 0.9  & 0.9337(14) & -0.38(2)
& 3.4 \\
 6 & 0.9300(48) & -0.46(7) & 0.51(16) & 1.1  & 0.9228(35) & -0.33(4) &
1.0 \\
 8 & 0.9144(41) & -0.38(5) & 0.38(11) & 0.4  & 0.9114(25) & -0.31(3) &
0.1 \\ \hline
\end{tabular}
\caption{\label{table_contTc} Continuum limits of $T_c/\surd\sigma$
for various $N$, using either a $O(a^4)$ or $O(a^2)$ formula,
with values of the coefficients of the lattice corrections, and the
quality of fit.}
\end{center}
\end{table}

\begin{table}
\begin{center}
\begin{tabular}{|c|c|c|}\hline
\multicolumn{3}{|c|}{latent heat } \\ \hline
$N$ & \rule[-3mm]{0mm}{8mm}$\frac{1}{T_c}\left(\frac{L_h}{N^2}\right)^\frac{1}{3}$ & $\chi^2/n_{df}$ \\  \hline
4  & 0.393(13) &  1.4 \\ 
5  & 0.448(4) &  1.4 \\ 
6  & 0.459(4) &  0.9 \\ 
8  & 0.453(4) &  0.9  \\   \hline
\end{tabular}
\caption{\label{table_LhTc} Continuum limit of the latent heat using
an $O(a^2)$ extrapolation to $2\leq L_t \leq 5$.}
\end{center}
\end{table}

\begin{table}
\begin{center}
\begin{tabular}{|cc|cccc|ccc|}\hline
\multicolumn{9}{|c|}{ $T_c/\surd\sigma$ vs $N$ } \\ \hline
\multicolumn{6}{|c|}{$\qquad \quad O(1/N^4)$ fit} & \multicolumn{3}{c|}{$O(1/N^2)$ fit}\\ \hline
  & $N \geq$ &  $N=\infty$ &  $c_1$ & $c_2$  & $\chi^2/n_{df}$  &
$N=\infty$ &  $c_1$ & $\chi^2/n_{df}$  \\ \hline
$a^4$ &  2 & 0.9035(41) & 0.85(11) & 0.12(31) & 0.23 & 0.9026(23) & 0.88(5)
& 0.18  \\ \hline
$a^2$ &  2 & 0.9016(26) & 0.76(7) & 0.46(33) & 0.70 & 0.8989(17) & 0.85(4)
& 0.98  \\ 
$a^2$ &  3 & 0.8977(38) & 0.94(14) & -1.0(1.1) & 0.07 & 0.9007(20) & 0.81(5)
& 0.35  \\ \hline
\end{tabular}
\caption{\label{table_NTcK} Results of fitting the $N$-dependence of
$T_c/\surd\sigma$ either to $O(1/N^4)$ or to $O(1/N^2)$. Results shown
separately for the different extrapolations to the continuum limit.} 
\end{center}
\end{table}

\clearpage

\begin	{figure}[p]
\begin	{center}
\leavevmode
\input	{su2Ns90Nt3beta4.9.tex}
\end	{center}
%\vskip 0.15in
\caption{Number of lattice fields, $N$, with a given
value of the Polyakov loop $|\overline{l}_p|$.
In SU(2), at $\beta=4.9$, on a $90^2 3$ lattice.}
\label{fig_2ndhisto}
\end 	{figure}

\begin	{figure}[p]
\begin	{center}
\leavevmode
\input	{betachisu2Nt3}
\end	{center}
%\vskip 0.15in
\caption{A plot of the critical coupling $\beta_c(V)$ against
$(L_s/L_t)^{-1/\nu}$ with an extrapolation to $V=\infty$.
In SU(2) for $aT_c=1/L_t=1/3$.}
\label{fig_bcVsu2}
\end 	{figure}

\begin	{figure}[p]
\begin	{center}
\leavevmode
\input	{betachisu6Nt3}
\end	{center}
%\vskip 0.15in
\caption{A plot of the critical coupling $y=\beta_c(V)$ against
$x=(L_s/L_t)^2$ with an extrapolation to $V=\infty$.
In SU(6) for $aT_c=1/L_t=1/3$.} 
\label{fig_bcVsu6}
\end 	{figure}

\begin	{figure}[p]
\begin	{center}
\leavevmode
\input	{SU4Ls60Lt5polyhist}
\end	{center}
%\vskip 0.15in
\caption{Histogram of the number of lattice fields with a given
value of the Polyakov loop $|\overline{l}_p|/N$.
In SU(4), at the $\beta$ indicated, on a $60^2 5$ lattice.}
\label{fig_histN4L60}
\end 	{figure}

\begin	{figure}[p]
\begin	{center}
\leavevmode
\input	{SU4Ls120Lt5polyhist}
\end	{center}
%\vskip 0.15in
\caption{Histogram of the number of lattice fields with a given
value of the Polyakov loop $|\overline{l}_p|/N$.
In SU(4), at the $\beta$ indicated, on a $120^2 5$ lattice.}
\label{fig_histN4L120}
\end 	{figure}

\begin	{figure}[p]
\begin	{center}
\leavevmode
\input	{betachisu4Nt3}
\end	{center}
%\vskip 0.15in
\caption{A plot of the critical coupling $y=\beta_c(V)$ against
$x=(L_s/L_t)^2$ with an extrapolation to $V=\infty$.
In SU(4) for $aT_c=1/L_t=1/3$.} 
\label{fig_bcVsu4}
\end 	{figure}

\end{document}